\documentclass[12pt,preprint]{aastex}
\newcommand{\gtsim}{\raisebox{-1.0ex}{$\stackrel{\textstyle>}{\sim}$}}
\newcommand{\ltsim}{\raisebox{-1.0ex}{$\stackrel{\textstyle<}{\sim}$}}
\def\kms{km~s$^{-1}$}
\def\al{Alfv\'{e}n}

\def\hinode{{\sl Hinode}}

\def\p78{{\sl P78-1}}

\def\iris{{\sl IRIS}}

\def\caii{Ca~{\sc ii}}

\def\siiv{Si~{\sc iv}}

\def\halpha{H$\alpha$}

\def\al{Alfv\'{e}n}

\def\kms{km~s$^{-1}$}

\def\etal{et~al.}

\begin{document}
%


\slugcomment{Published 2016 Sep 1: \underline{ApJ {\it Letters}}, 828, L9}

\title{A Microfilament-Eruption Mechanism for Solar Spicules}

\author{Alphonse C.~Sterling\altaffilmark{1} \& Ronald L. Moore\altaffilmark{1,2}}

\altaffiltext{1}{ Heliophysics and Planetary Science Office, ZP13, Marshall Space Flight Center, Huntsville, AL 35812, USA;
alphonse.sterling@nasa.gov, ron.moore@nasa.gov}

\altaffiltext{2}{Center for Space Plasma and Aeronomic Research, University of Alabama in Huntsville, 
Huntsville, AL 35899, USA}

\begin{abstract}

Recent investigations indicate that solar coronal jets result from eruptions of small-scale chromospheric
filaments, called minifilaments; that is, the jets are produced by scaled-down versions of typical-sized
filament eruptions.  We consider whether solar spicules might in turn be scaled-down versions of coronal
jets, being driven by eruptions of {\it microfilaments.} Assuming a microfilament's size is about a
spicule's width ($\sim$300~km), the estimated occurrence number plotted against the estimated size of
erupting filaments, minifilaments, and microfilaments approximately follows a power-law distribution 
(based on counts of CMEs, coronal jets, and spicules),
suggesting that many or most spicules could result from microfilament eruptions. Observed spicule-base
\caii\ brightenings plausibly result from such  microfilament eruptions.  By analogy with coronal jets,
microfilament eruptions might produce spicules with many of their observed characteristics, including
smooth rise profiles, twisting motions, and EUV counterparts.  The postulated microfilament eruptions
are presumably eruptions of twisted-core micro magnetic bipoles that are $\sim$1$''.0$ wide. These
explosive bipoles might be built and destabilized by  merging and cancelation of magnetic-flux elements
of $\sim$few$\times 100$~G and of size $\ltsim 0''.5$---$1''.0$.  If however spicules are relatively more
numerous than indicated by our extrapolated distribution, then only a fraction of spicules might result
from this proposed mechanism.


\end{abstract}


\section{Introduction}
\label{sec-introduction}

Ground-based observation of solar spicules has a long history \citep[e.g.][]{beckers68}.  For $\sim$10
years now they have been observed from space with the \caii\ filter of the Solar Optical Telescope (SOT)
on \hinode. Viewed at the limb with SOT, \citet{depontieu_et07} report that spicules observed in 
quiet-Sun and coronal-hole regions have different average properties from those typically found in 
active regions, calling the former type~II spicules and the latter type~I spicules.  Among the 
differences, type~II spicules have velocities and lifetimes of $\sim$30---110~\kms\ and $\gtsim$50---150~s,
respectively, while type~I spicules have velocities and lifetimes of
$\sim$15---40~\kms\ and $\sim$150---400~s \citep{depontieu_et07,pereira_et12,skogsrud_et15}.

(A side note: Discussions are ongoing on whether there is an essential difference between
spicule ``types''; see \citeauthor{zhang_et12}~(\citeyear{zhang_et12}) and 
\citeauthor{pereira_et12}~(\citeyear{pereira_et12}), and also
\citeauthor{skogsrud_et15}~(\citeyear{skogsrud_et15}).  It is still uncertain how the historical
ground-based-observed spicules described by, e.g., \citeauthor{beckers68} (usually observed in
\halpha) correspond to type~I and type~II spicules (usually observed by \hinode\ in \caii); following 
\citeauthor{sterling_et10a}~(\citeyear{sterling_et10a}) and
\citeauthor{pereira_et13}~(\citeyear{pereira_et13}), we call the historical ground-based-observed
features ``classical spicules.'' We do not belabor questions about spicule types here, but expect
clarification  from future studies.)

What drives solar spicules?  \citet{depontieu_et04} suggest that active region
(type~I) spicules result from waves generated by photospheric p-modes.  The
mechanism for the much-faster-moving type~II spicules however is seemingly a greater
mystery: p-mode waves move through the chromosphere at roughly the acoustic sound
speed ($\sim$10~\kms), and in non-linear simulations they can result in features
that look like type~I spicules (with velocities $\sim$20~\kms), but not the
much-faster ($\sim$100~\kms) type~II spicules.  Detailed MHD modeling by
\citet{martinez-sykora_et13} resulted in a simulated feature that resembled a
type~II spicule, but one of several difficulties noted by the authors is that
this mechanism does not produce type~II spicules with nearly-enough frequency to
match the huge numbers of observed spicules.  Other recent models do
not produce 100~\kms\ spicules \citep[e.g.][]{iijima_et15}, and older spicule models
also have difficulties \citep{sterling00}.

On a size scale substantially larger than spicules, jet-like structures also are
commonly observed with coronal imaging instruments, in X-rays and EUV
\citep[e.g.][]{shibata_et92,cirtain_et07,nistico09}.  A
natural question is whether these coronal jets might be larger-scale versions of
spicules.  Recent observations indicate that coronal jets result from eruptions of
small-scale filaments, or {\it minifilaments} \citep{sterling_et15,sterling_et16}. 
Here we explore the idea of whether by the same mechanism spicules may result from eruptions of
filament-like objects smaller than minifilaments, which we term {\it microfilaments.}

\section{Coronal Jets}
\label{sec-coronal jets}

Images from the \hinode/X-ray telescope (XRT) show X-ray jets in coronal holes 
occurring at a rate $\sim$60/day, with average widths of 8000~km and lifetimes of
$\sim$10~min \citep{savcheva_et07}. \citep[See, e.g.][for active-region-jet
parameters.]{shimojo_et96}  Many, if not all, X-ray jets have EUV-jet counterparts
\citep[e.g.][]{raouafi_et08,moore_et13,sterling_et15}.  Various studies also show that
X-ray jets usually have a brightening off to one side of their base; we refer to that
brightening as the ``jet bright point'' (JBP)\@. \citet{sterling_et15} found that a
jet minifilament erupts from the location where the JBP forms.  They concluded that
polar-coronal-hole coronal jets are scaled-down versions of typical-sized filament
eruptions that drive a coronal mass ejection (CME) and produce a solar flare at the
site of the pre-eruption filament, where analogously the minifilament eruption drives the
jet and produces a miniflare (the JBP) at the site of the pre-eruption minifilament. 
\citet{sterling_et16} found that at least some active-region coronal jets also result 
from minifilament eruptions. Figure~1 shows their schematic interpretation of the mechanism 
producing the jets.

What leads to the minifilament eruptions has not yet been addressed systematically. 
Several on-disk studies of individual jets suggest that nearly-concurrent flux
cancelation frequently occurs at the base
\citep[e.g.,][]{huang_et12}.  In the neighborhood of
active regions, there is evidence that sometimes emerging flux together with canceling flux lead
to the minifilament eruptions \citep{chandrashekhar_et14,sterling_et16}.

\section{Spicules as Small-Scale Coronal Jets?}
\label{sec-spicule jets}

Just as CMEs and coronal jets respectively are consequences of filament eruptions and 
minifilament eruptions, perhaps spicules are consequences of even-smaller-scale 
microfilament eruptions.  

To consider this suggestion's plausibility, we estimate the size-scale distribution of erupting
filament-like objects from the number of eruptions of three size classes: eruptions of typical
filaments, minifilaments, and (if they exist) microfilaments, respectively corresponding to CME/flare
eruptions, coronal jets, and spicules.


For typical-filament lengths, \citet{bernasconi_et05} gives a range of $3\times 10^4$---$1.1\times 10^5$~km;
we will use $(7\pm 4) \times 10^4$~km.  For minifilaments, \citet{sterling_et15} found sizes of $(8 \pm 3)
\times 10^3$~km.  Next we must estimate what size a postulated spicule  microfilament might be.  We note that the
average measured size of 8000~km for the length of coronal-jet minifilaments from Sterling~\etal~(2015) is the
same as the average width of X-ray jets measured by \citet{savcheva_et07}.  Similarly we hypothesize that the
spicule-microfilament length would be about equal to the spicule width, which is $\sim$300~km
\citep{pereira_et12}; we assume a range $\pm 100$~km.

How many of each class of filament eruption at a time might we expect?  A value is readily available for
the number of spicules  (reaching above 3000~km) on the Sun at any time: that is  $9.3 \times 10^4$ from
\citet{athay59}. \citet{lynch_et73} suggest numbers of $\ltsim 5$-times higher; we will use this as our
upper limit.  We do not know a similar quote for the number of coronal jets on the Sun at any given time,
but we estimate this using the value of 60 X-ray jets/day in the two polar coronal holes from
\citet{savcheva_et07}. Averaged over a day, this is  2.5 jets/hour, or 0.42 jets every 10 min.   Since the
average lifetime of a jet is 10~min \citep{savcheva_et07}, we expect $\sim$0.5 jet among the two polar
coronal holes at any given time (i.e., looking at two independent times will likely detect one jet on
average).  At the time of the  \citet{savcheva_et07} study, each of the two polar coronal holes covered
$\ltsim$5\% of the Sun's surface \citep{hess_webber_et14}.  So we might guess that there would be about
ten times the number of jets over the whole Sun as were in the polar coronal holes. (Coronal jets are
common in active regions and quiet Sun, e.g., \citeauthor{shimojo_et96}~\citeyear{shimojo_et96}, in
addition to coronal holes.) So we estimate that there are $\sim$5 coronal jets on the Sun at any given
time.  For ``typical'' filament eruptions, we note that CMEs occur at between 0.5 and 6 per day
\citep{yashiro_et04,chen11}; using a C/M-class flare's strong-phase duration (NOAA-listed
end-minus-start time) of $\sim$20~min \citep{veronig_et02} to estimate a CME's filament-eruption-phase
duration gives $\sim$0.03 CME-producing typical filament eruptions occurring on the Sun at any given
time.

Figure~2 plots these values, giving a number-versus-size distribution for filament-eruption-like events
on the Sun, going down to microfilament eruptions postulated to produce spicules.  This plot indicates 
that the idea of a microfilament-eruption origin of spicules is
consistent with a power-law distribution of filament sizes from large-scale erupting filaments of typical
solar eruptions, through erupting coronal-jet minifilaments and on down to spicule-producing erupting
microfilaments.


If spicules are more numerous than the above-cited measurements determined, as perhaps indicated by
\hinode/SOT observations, then our extrapolated distribution might substantially underestimate the
number of spicules.  In that  case only a fraction of spicules may result from erupting microfilaments. 
In at least some cases however, the \hinode/SOT observations show many narrow spicule
strands within a wider-erupting feature, giving an overcount of spicule-producing eruptions 
\citep[][]{sterling_et10a}. Further studies should address this question.

\section{Consequences and Expectations for Microfilament-Eruption Spicules}
\label{sec-consequences}

If microfilament eruptions drive spicules, we would expect a brightening corresponding to the
JBP observed in solar jets (Fig.~1).  Earlier numerical-simulation spicule models
predicted brightening at the base of spicules near the time of their formation, but 
lack of identification of such brightenings in \halpha\
\citep{suematsu_et95} has been presented as evidence against such models
\citep{sterling00}.  More recently however, subtle brightenings at the base of some rising
spicules {\it have} been observed by \cite{sterling_et10b} in near-limb \hinode/SOT
\caii\ movies.  They found a plethora of small ($\ltsim 0''.5$), transient (few-100~s
lifetime) ``\caii\ brightenings'' moving laterally at $\sim$10~\kms, some of which occurred when
and where spicules formed.  Figures~3(a---c)
and~3(d---f) show two other such SOT~\caii\ episodes, processed
only with ``fg\_prep" and an unsharp mask.  On-disk spicules are difficult to see in SOT/\caii\ images 
\citep{beck_et13}, and only a few are identifiable just
inside the limb; some brightenings may be source locations for some invisible-against-the-disk 
spicules.  In Figure~3(g---i) we show a
\hinode/XRT X-ray jet from \citet{sterling_et15}, with a color-reversed scaling. 
Except for the size-scale difference, coronal-jet morphology in the X-ray images is
similar to the spicule morphology in the SOT images, in terms of the jet/spicule base 
brightenings.  (Non-color-reversed XRT and AIA images and movies of the Figure~3(g---i)
jet are available in \citeauthor{sterling_et15}~\citeyear{sterling_et15}.)   Thus it
is plausible that some base brightenings and accompanying spicules in the SOT
\caii\ movies may be due to microfilament eruptions, just as the JBP brightenings and
coronal jets in the XRT and AIA movies are due to minifilament eruptions.
(See, e.g., \citeauthor{rutten_et91}~\citeyear{rutten_et91} and \citeauthor{sterling_et10b}~\citeyear{sterling_et10b} 
for possible non-microfilament-eruption causes for \caii~brightenings.)

Transient brightenings and small-scale jets are also seen at transition region temperatures by the
Interface Region Imaging Spectrograph (\iris) spacecraft \citep{tian_et14}, and base brightenings  of
small-scale active region \hinode/\caii-observed ``Ca jets'' \citep{shibata_et07} are also candidates for
base brightenings from putative microfilament eruptions.

There is evidence that coronal jets consist of two components: a hotter component, visible in X-rays and
in hotter AIA EUV channels such as 171~\AA, 193~\AA, and 211~\AA; and a cooler component frequently
visible in cooler AIA EUV channels, especially 304~\AA\ \citep{moore_et13,sterling_et15}.  In the
erupting-minifilament picture we expect the hotter component to form when the outer envelope of the
exploding minifilament field (the part that does not contain minifilament material) undergoes
reconnection with the ambient coronal field (``envelope reconnection''; Fig.~1(b)), and the cooler jet 
to form when the core of the exploding field (the part containing the cool material) undergoes that 
reconnection (``core reconnection''; Fig.~1(c)).  Similar core and envelope reconnections should occur 
if microfilament eruptions make spicules.  In that case, analogously, the core reconnection would 
expel upward cool chromospheric material to form the commonly-observed chromospheric spicules, while 
the envelope reconnection could produce the EUV spicules reported by \citet{depontieu_et11} (and perhaps 
also UV spicules); the EUV spicules would be a co-produced hotter component of 
chromospheric spicules, and would not necessarily imply that spicules supply a large portion of the corona's 
hot material \citep{madjarska_et11,klimchuk_et14}.  Additionally, the microfilaments might either 
fully erupt, or be partially confined, as with minifilaments \citep{sterling_et15}.

\section{Summary and Discussion}
\label{sec-discussion}

Recent observations indicate that coronal
jets result from minifilament eruptions, and so a natural suggestion is that spicules result from
eruptions of even smaller ``microfilaments''  \citep[cf.][]{moore90}.  \citet{moore_et11} considered that
spicules are small-scale jets, but that was based on earlier theoretical ideas of how coronal jets worked
\citep[e.g.][]{shibata_et92}, prior to our inferred minifilament scenario of Figure~1
\citep{sterling_et15}.  Based on that new picture, our rough estimation of the number distribution of
filament-eruption events (Fig.~2) allows that the majority of spicules could operate in this fashion.  


A microfilament-eruption mechanism could address several hitherto-puzzling aspects of spicules:

\begin{itemize}

\item Spicule energy source. Non-potential magnetic energy stored in the core of the
pre-eruption microfilament field would power the spicules.

\item Smooth rise trajectories.  \hinode\ observations indicate that spicules accelerate smoothly 
\citep[e.g.][]{depontieu_et07}, which is different from the predicted velocity jumps
of some earlier models \citep[see][]{sterling00}.  Cool coronal jets (likely due
to minifilament eruptions) observed in AIA 304~\AA\ images also follow smooth trajectories 
\citep{moschou_et13}. So analogously we would expect microfilament eruptions to result in spicules 
following smooth trajectories, as observed (although detailed comparisons of the respective trajectories have 
not yet been made).

\item Twisting spicules.  Long-suspected twisting 
of spicules have now been observed with confidence \citep[e.g.][]{depontieu_et12}. 
Twisting is also observed in coronal jets, including ones clearly due to minifilament
eruptions \citep[][and references therein]{moore_et15}.  This twist could be conveyed
from the erupting closed, twisted field carrying the minifilament to open ambient field
by the external reconnection of Figure~1(b), and~1(c), as in \citet{shibata_et86}.  A
similar process could explain twisting of spicules if they result from microfilament
eruptions.

\item Spicule lateral motions.  \citet{depontieu_et12} and other workers note
transverse motions of spicules.  Lateral (transverse) motions also occur on X-ray jets
\citep{cirtain_et07,savcheva_et07}.  A jet from an erupting micro- or minifilament
might acquire undulatory horizontal motion from the reconnection with the ambient field; cf.\
Fig.~1(b). Also, this reconnection could produce progressive illumination of
the ambient field lines, resulting in apparent splitting \citep{sterling_et10b}.

\item Hotter spicule components. EUV and UV spicules (\S\ref{sec-consequences} above, and 
references in \citeauthor{sterling00}~\citeyear{sterling00})
could result naturally when the envelope of the microfilament field carrying less-dense
material reconnects with the ambient field (\S\ref{sec-consequences}).

\item Gap at spicule bases.  The sometimes-observed gap between the photospheric limb and 
the bottoms of spicules \citep[e.g.,][]{gaizauskas84} could
result from the spire of the spicule forming by reconnection at the height of the
neighboring bipole (Fig.~1).

\end{itemize}

We can estimate the pre-eruption magnetic field strength, $B$, required for the postulated
microfilament flux ropes \citep[cf.\@][]{falconer_et03} that erupt to produce spicules. 
\citet{sterling00} estimates a spicule to have energy $\ltsim$10$^{25}$~erg. (Dimensions for a
type~II spicule from, e.g., \citeauthor{pereira_et12}~\citeyear{pereira_et12}, are smaller than those
assumed in \citeauthor{sterling00}~\citeyear{sterling00}, implying lower energy.  But their type~II
spicule speeds are larger than assumed in \citeauthor{sterling00}~\citeyear{sterling00}, and \al\
wave energy was not included in \citeauthor{sterling00}~\citeyear{sterling00} and could add
$\sim$50\% to the energy (\citeauthor{moore_et11}~\citeyear{moore_et11}).  So 
$\sim$10$^{25}$~erg is a reasonable estimate.)  We equate this to the flux-rope total magnetic energy
of $V(B^2/8\pi)$, where the flux rope of length $l$ and radius $r$ has volume $V$ of $\pi r^2l$. 
Taking $l\sim$500~km (about the size of the \caii\ brightenings in Fig.~3, and similar to spicule
width used in \citeauthor{sterling00}~\citeyear{sterling00}), and estimating $l/r\sim 10$ \citep[as
in filaments; cf.\@][]{falconer_et03}, gives $B\sim 250$~G\@ for the flux rope, and hence for the
host bipole.  Network-boundary magnetic cancelations of this size are observed 
\citep[e.g.][]{gosic_et16}, but it has not yet been demonstrated that the strength of the canceling 
field is of the correct magnitude or that the canceling field is connected to spicules.


We can estimate the temperature regime in which the base brightenings should appear by equating the
estimated spicule  energy, $E,$ deposited in a volume, $V,$ as thermal  energy, $(3/2) n V k T$, where
$n$, $k$, and $T$ are particle number density, Boltzmann's constant, and
temperature, respectively.  With $V$ as above, this gives $E \sim [(3/2) 10^{-2}\pi k] n l^3 T.$ For a
spicule/microfilament, using $E$ and $l$ as above, and $n \sim 10^{14}$~cm$^{-3}$ \citep[][model~C at
755 km]{VAL}, gives $T \sim 10^5$~K, consistent with chromospheric/transition region emission (e.g., \halpha, 
\siiv\@). For a
jet/minifilament, taking  $E \sim 10^{27}$~erg \citep{pucci_et13}, $l=8000$~km \citep{sterling_et15},
and $n \sim 10^{11}$  \citep[][model~C at 1990 km]{VAL}, gives $T \gtsim 10^6$~K, consistent with
coronal emission.  Modeling is required to test in detail predictions of the 
proposed mechanism.


Theoretical calculations suggest that type~I spicules might result from p-mode oscillations 
\citep[e.g.,][]{hansteen_et06}.  Erupting microfilaments also might occur in active regions  too,
leading to type~I spicules. Moreover, p-mode oscillations could conceivably drive repeated 
cancelations,  each leading to an erupting microfilament flux rope (perhaps driving waves/shocks). 
This is still speculation, however.

Our conjectured microfilaments may have not yet been detected because of
their expected small size ($\ltsim$0.$''5$---$1''.0$).  In addition, \citet{sterling_et16}
suggested that minifilaments that cause jets in active regions might often not be
visible when low to the photosphere because of surrounding obscuring material. 
Similarly, spicule-producing microfilaments might be unobscured, and hence observable, only infrequently.
Finally, the microfilament-carrying twisted-core magnetic bipole might  only be created
near or during the microfilament-eruption time, making the pre-eruption
microfilament short lived.  Further observations may provide additional hints as to whether 
microfilament eruptions make spicules, but a clear determination may have to await a new generation 
of instruments, such as DKIST or the next space-based follow on to \hinode/SOT\@.



\acknowledgments

This work was supported by funding from the Heliophysics Division of NASA's Science Mission Directorate through
the Heliophysics Guest Investigator (HGI)  Program, and the \hinode\ Project. We thank T. Tarbell for assistance
with SOT images.  \hinode\ is a Japanese mission developed and launched by ISAS/JAXA, with NAOJ as domestic
partner and NASA and STFC (UK) as international partners, and operated  by these agencies in co-operation with
ESA and NSC (Norway).  Figure~1, and Figure and video 3(g---i), adapted by permission  from Macmillan Publishers
Ltd: Nature, \citet{sterling_et15}, copyright 2015.

\clearpage

\begin{figure}
\hspace*{-1.3cm}\includegraphics[angle=-90,scale=0.70]{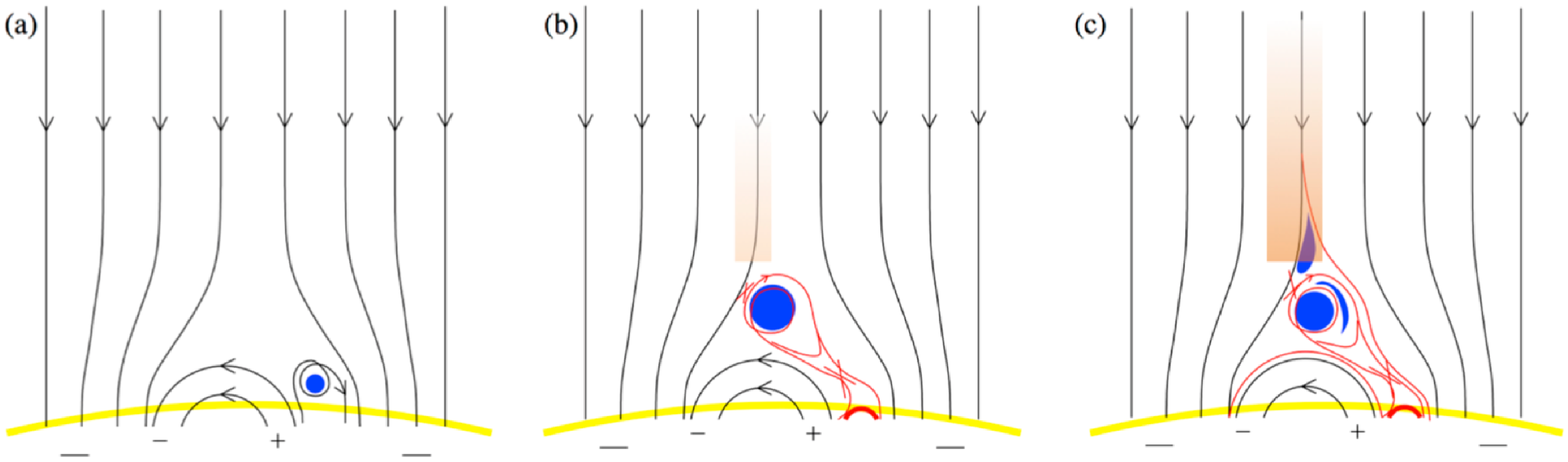}
\caption{\small Schematic of minifilament eruption leading to a coronal jet (modified version of
\citeauthor{sterling_et15}~\citeyear{sterling_et15}, Fig.~2).  Black lines show magnetic fields,
with arrows indicating polarities. Red lines indicate fields that have undergone magnetic
reconnection, with red ``X''-es showing reconnection locations. (a) Initially, a minifilament (blue)
sits in a compact sheared bipole adjacent to a larger less-sheared bipole, all inside surrounding
open field.  (b) An unspecified agent triggers the minifilament field to erupt like a larger-scale 
filament eruption.  Reconnection
at the left-side red X between the envelope of the erupting minifilament field and the open coronal field results in
a hot jet (shaded-orange strip), visible as a hotter-EUV (e.g., 211~\AA) or X-ray jet.  Reconnection
interior to the exploding minifilament field at the right-side red X results in the JBP (bold red
semicircle), analogous to a solar flare in large-scale filament eruptions. (c) If the minifilament
erupts far enough into the opposite-polarity open-field region, then the left-side reconnection
``eats away'' enough of the enveloping outer field surrounding the minifilament in the core for
minifilament material to be expelled along the vertical field, resulting in a cool (e.g., EUV 304~\AA) jet component. 
See \citet{sterling_et15,sterling_et16} for details.}
\end{figure}
\clearpage

\begin{figure}
\epsscale{0.9}
\plotone{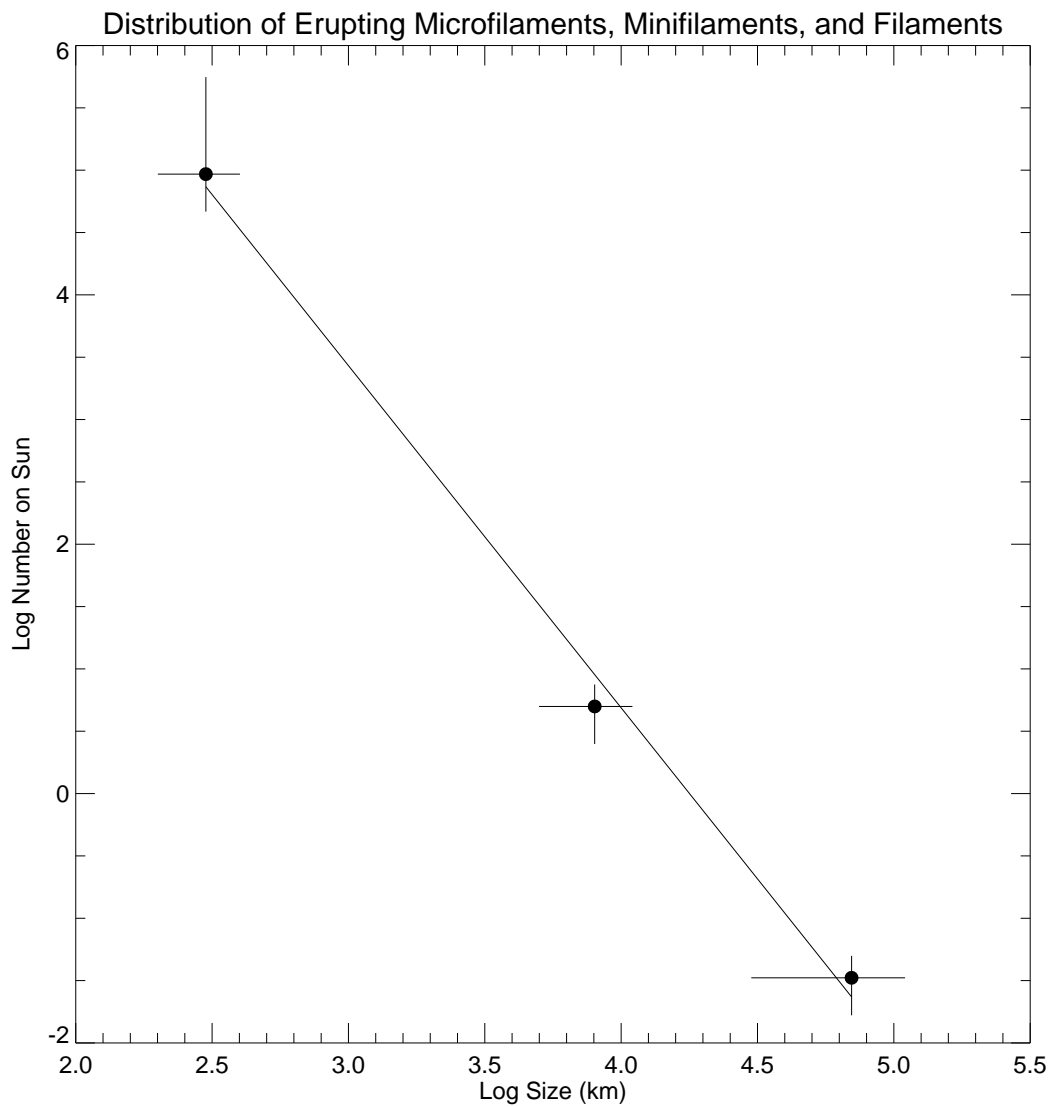}
\caption{\small Distribution of estimated number of erupting-filament-like features on the Sun at any given
time, as a function of the size of those erupting features.  Right, middle, and left points are for
filament eruptions driving CMEs and typical solar flares,  minifilament eruptions driving coronal jets
and JBPs, and postulated {\it microfilament} eruptions that would drive spicules. ``Error'' bars show
measured or estimated ranges of the plotted values.  We used $\pm$50\% of the plotted values as guesses
for the uncertainties in the number-on-Sun values, except the upper number for the erupting
microfilaments is from \citet{lynch_et73}. See text for uncertainties in sizes. The line ($\rm{slope} =
-2.7$) is the least-squares best fit to the three points without consideration of uncertainty weights.}

\end{figure}
\clearpage

\begin{figure} 
\hspace*{0.5cm}\includegraphics[angle=270,scale=0.70]{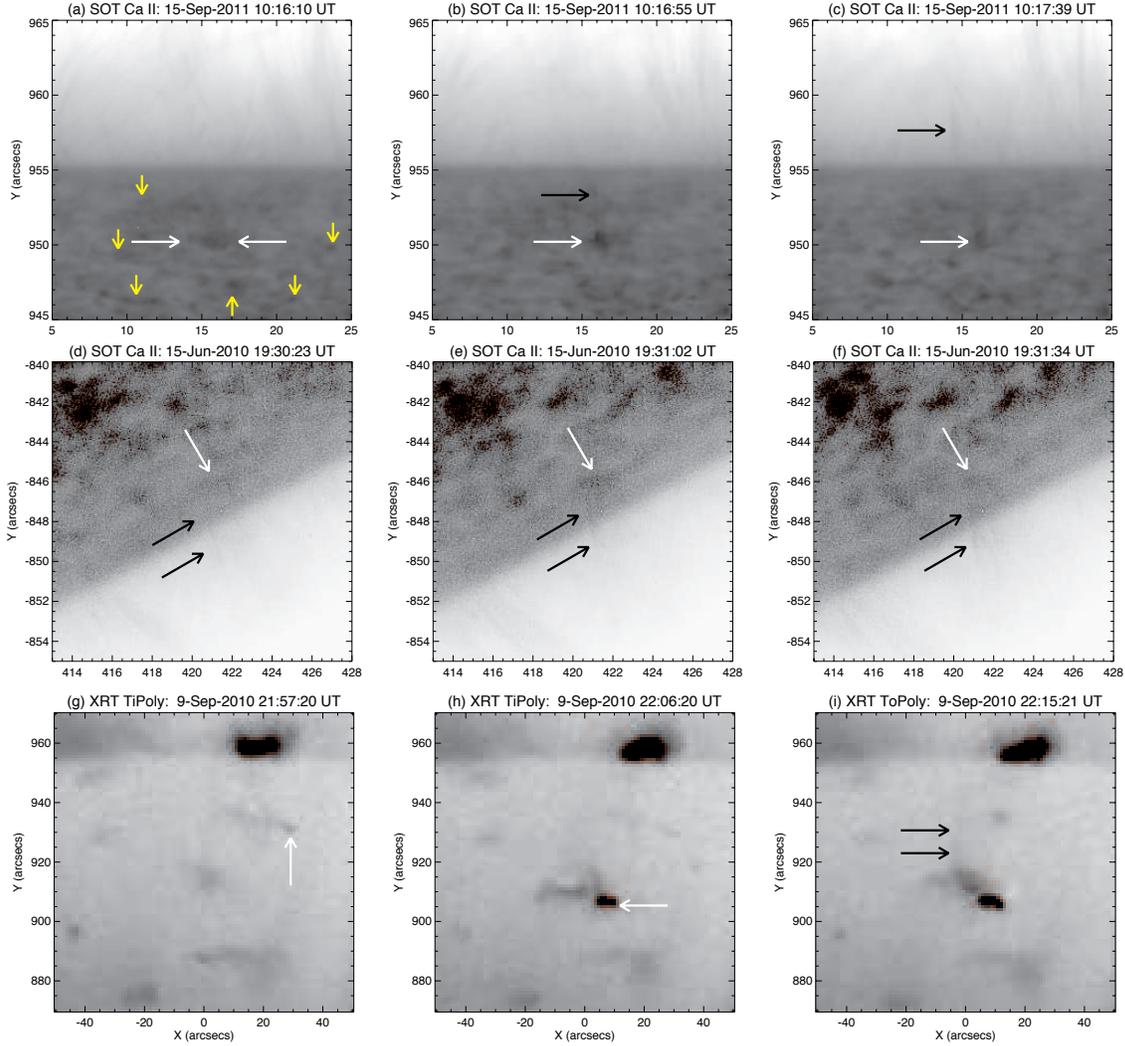}
\caption{\small (a---c) \hinode\ SOT \caii\ color-reversed images ($\sim$0$''.1$/pixel resolution) of a
polar coronal hole near the limb, showing on-disk ``\caii\ brightenings'' and spicules (white/black 
arrows show brightenings/spicules). Yellow arrows show show examples of additional brightenings that 
might also be at roots
of spicules.) Brightenings between white arrows in
(a) evolve to a more-concentrated brightening indicated by arrow in (b), that fades
significantly by (c).  Black arrows in (b) and (c) show a strand of a Type~II spicule emanating from
the \caii\ brightening.  (d--f) Same as a---c, but for a quiet-Sun region near the south pole limb. 
Arrows in accompanying videos show additional base 
brightening/spicule examples.  (g---i) Images from \hinode/XRT, showing X-ray jet \#12 in
\citet{sterling_et15} (Extended Data Fig.~2 of that paper); here we show the same movie as in that
paper, but with colors reversed to emphasize morphological similarities with the SOT features in
(a---f); white/black arrows show JBP examples and a jet spire, respectively, where the JBP in (g) is 
of a weaker jet in the background.  North is upward and west is right.  Intensity 
scaled to highlight faint features, resulting in saturation of some
features.}

\end{figure}

\end{document}